
\input phyzzx.tex
\input epsf
\overfullrule = 0pt
\def\ie{{\it i.e.~}}

\def\oh{{\textstyle {1\over 2}}}

\def\text{\textstyle}

\def\N{{$1\over N$}~}

\def\refmark#1{[#1]}                        
\def\Q{{${\rm QCD}_2$}~}
\def\2d{{two dimensions}~}
\def\sp{\,\,\,\,}
\def\m{{\mu}}
\def\n{{\nu}}

\magnification=\magstep1
\vsize=23.5 true cm
\hsize=15.4 true cm
\predisplaypenalty=0
\abovedisplayskip=3mm plus6pt minus 4pt
\belowdisplayskip=3mm plus6pt minus 4pt
\abovedisplayshortskip=0mm plus6pt
\belowdisplayshortskip=2mm plus6pt minus 4pt
\normalbaselineskip=12pt

\REF\thooft{G. 't Hooft, {\sl Nucl. Phys. }{\bf B72}, 461 (1974).  }
\REF\kazkos{ D. Weingarten, {\sl   Phys. Lett. } {\bf B90}, 285 (1990); V.
Kazakov and I. Kostov, {\sl   Phys. Lett. } {\bf B128}, 316 (1983); I. Kostov
{\sl Phys. Lett.} {\bf B138}, 191 (1984); {\sl Nucl. Phys. } {\bf B179}, 283
(1981); K. OBrien and J. Zuber, {\sl Nucl. Phys. }{\bf B253}, 621 (1985) }
\REF\growit{D. Gross and E. Witten, {\sl Phys. Rev.} {\bf D21}, 446(1980). }
\REF\polch{J. Polchinski,  Texas preprint, UTTG-26-91 (1991) }
\REF\gromen{D. Gross and P. Mende, {\sl Physics Letters \bf 197B}, 129 (1987).}
\REF\spec{J. Polchinski, {\sl Phys. Rev. Lett} {\bf 68}, 1267 (1992);  M.
Green, QMW-91-24 (1991)}
\REF\kazkostwo{ V. Kazakov and I. Kostov, {\sl    Nucl. Phys.} {\bf B220}, 167
(1983); I. Kostov {\sl Nucl. Phys. Lett.} {\bf B265}, 223 (1986).}
\REF\calgro{C. Callan, N. Coote and D. Gross, {\sl Phys. Rev. \bf D13}, 1649
(1976)}
\REF\gropol{D. Gross and A. Polyakov, unpublished}
\REF\pol{ A. Polyakov, {\sl    Nucl. Phys.} {\bf B40}, 235 (1982).}
\REF\migdal{A. Migdal, {\sl Zh. Eksp. Teor. Fiz.} {\bf 69}, 810 (1975)( Sov.
Phys. Jetp. {\bf 42}, 413).}
\REF\rus{B. Rusakov, {\sl Mod. Phys. Lett. } {\bf A5}, 693 (1990).}
\REF\witetal{ E. Witten, {\sl Comm. Math. Phys}  {\bf 141}, 153(1991; D.Fine,
{\sl Comm. Math. Phys.} {\bf 134}, 273 (1990); M. Blau and G. Thompson,
NIKHEF-H/91-09, MZ-TH/91-17. }
\REF\knesser{ H. Kneser, {\sl  Math. Ann.} {\bf   103}, 347 (1930); H.
Zieschang, E. Vogt and H. Coldewey, {\sl Lecture Notes in Mathematics,{\bf Vol.
835}}, Springer-Verlag, New York, 1980.}
\REF\tobewat{ D. Gross and W. Taylor, to be published}

\Pubnum{ LBL 3323? \cr  PUPT 1356 }
\date{December, 1992 }

\titlepage
\vsize=23.5 true cm
\hsize=15.4 true cm
\title{ {\bf TWO DIMENSIONAL QCD AS A STRING THEORY}}
\foot{This work was supported in part by the Director, Office of
Energy Research, Office of High Energy and Nuclear Physics, Division of
High Energy Physics of the U.S. Department of Energy under Contract
DE-AC03-76SF00098 and in part by the National Science Foundation under
grant PHY90-21984.}

\author{DAVID J. GROSS\
\foot{ On leave from Princeton University, Princeton, New Jersey.} }
\address{   Theoretical Physics Group \break
Physics Division \break
Lawrence Berkeley Laboratory    \break
1 Cyclotron Road \break
Berkeley, California 94720  }
\vskip 1truein

\abstract{  I explore the possibility of finding an equivalent string
representation of two dimensional QCD. I  develop the large N expansion of the
\Q partition function on an arbitrary two dimensional Euclidean
manifold. If this  is related to a two-dimensional string theory then many of
the coefficients of the \N expansion must vanish. This is shown to be true to
all orders, giving strong evidence for the existence of a string
representation.}

\endpage

\vskip 1in

\centerline{\bf Disclaimer}

\vskip .2in

This document was prepared as an account of work sponsored by the United
States Government.  Neither the United States Government nor any agency
thereof, nor The Regents of the University of California, nor any of their
employees, makes any warranty, express or implied, or assumes any legal
liability or responsibility for the accuracy, completeness, or usefulness
of any information, apparatus, product, or process disclosed, or represents
that its use would not infringe privately owned rights.  Reference herein
to any specific commercial products process, or service by its trade name,
trademark, manufacturer, or otherwise, does not necessarily constitute or
imply its endorsement, recommendation, or favoring by the United States
Government or any agency thereof, or The Regents of the University of
California.  The views and opinions of authors expressed herein do not
necessarily state or reflect those of the United States Government or any
agency thereof of The Regents of the University of California and shall
not be used for advertising or product endorsement purposes.

\vskip 2in

\centerline{\it Lawrence Berkeley Laboratory is an equal opportunity employer.}

\endpage

\chapter{\bf Introduction}

 	It is an old idea that QCD might be represented as a string theory. This
notion dates back even before the development of QCD. Indeed, string
theory itself was stumbled on in an attempt to guess  simple
mathematical representations of strong interaction scattering
amplitudes which embodied some of the features gleamed from the
experiments of the 1960's. Many of the properties of hadrons are
understandable if we picture the hadrons as string-like flux tubes. This
picture is consistent with linear confinement, with the remarkably
linear Regge trajectories  and with the
approximate duality of hadronic scattering amplitudes.

	Within QCD itself there is internal  support for this idea. First, the \N
expansion of weak coupling perturbation theory can be
interpreted as corresponding to an expansion of an equivalent string
theory in which the string coupling is given by \N. This is the famous
result of 't-Hooft's analysis of the \N expansion of perturbative
QCD\refmark\thooft . The same is true for any {\em matrix model}--\ie a model
invariant under $SU(N)$ or $U(N)$, in which the basic dynamical variable is a
matrix in the adjoint representation of the group. The Feynman graphs in such a
theory can be represented as triangulations of a two dimensional surface.
't-Hooft's principal result was that one can use \N to pick out the topology,
\ie the genus=number of handles, of the surface,
since a diagram which corresponds to a genus $G$ Riemann surface is
weighted by $({1\over N})^{2G-2}$. The leading order in  the expansion of the
free energy in powers of \N is given by the planar graphs and is proportional
to $N^2$.

Another bit of evidence comes from the strong coupling lattice
formulation of the theory. The  strong coupling expansion
of the free energy can indeed be represented as a sum over surfaces
\refmark\kazkos .
Again there is  a natural large $N$ expansion which picks out definite
topologies for these surfaces. This result  is an existence proof for a
string formulation of QCD. However, the weights of the surfaces are extremely
complicated and it is not at all clear how to take the
continuum limit.\foot{ There is also the problem that for large $N$ there
is typically a phase transition between the strong and weak coupling
regimes \refmark\growit .}

{}From quite general considerations we expect that the large $N$ limit of
QCD is quite smooth, and should exhibit almost all of the qualitative
features of theory. Thus an expansion in powers of ${1\over 3}$ or
$({1\over 3})^2$ might be quite good. The longstanding hope has been
to find an equivalent (dual) description of QCD as some kind of string
theory, which would be useful in to calculate properties of the theory
in the infrared. QCD is a permanent part of the theoretical landscape and
eventually we will  have to develop analytic tools for dealing with the
theory in the infra-red. Lattice techniques are useful but they have not yet
lived up to their promise. Even if one manages to derive the hadronic
spectrum numerically, to an accuracy of 10\% or even 1\%, we will not be truly
satisfied unless we have some analytic understanding of the results.
Also, lattice Monte-Carlo methods can only be used to answer a small set of
questions. Many issues of great conceptual and practical interest--in
particular the calculation of scattering amplitudes, are thus far beyond
lattice control.
Any progress in controlling QCD in an explicit analytic fashion would be of
great conceptual value. It would also be of great practical aid to
experimentalists, who must use rather ad-hoc and primitive models of QCD
scattering amplitudes to estimate the backgrounds to interesting new physics.

The problems with this approach are many. First, if QCD is describable as a
string theory it is not as simple a theory as that employed for critical
strings. It appears to be easier to guess the string theory of everything than
to guess the string theory of QCD.
Most likely the weights of the surfaces that one would have to sum over will
depend on the  {\em extrinsic geometry} of the surface and not only its
intrinsic geometry. We know very little about such string theories.
Also there are reasons to believe that a string formulation would
require many (perhaps an infinite) new degrees of freedom in
addition to the coordinates of the string \refmark\polch . Finally, there is
the important conceptual problem--how do strings manage to look like particles
at short distances. The one thing we know for sure about QCD is that at large
momentum transfer hadronic scattering amplitudes have
canonical powerlike behavior in the momenta, up to calculable
logarithmic corrections. String scattering, on the other hand, is
remarkably soft. Critical string scattering amplitudes have, for
large momentum transfer, Gaussian fall-off \refmark\gromen.  How do QCD strings
avoid
this?  Recently there have been some interesting speculations regarding this
problem  \refmark\spec , however it remains open.

In this paper I shall  present some evidence in favor of the hypothesis that
two-dimensional QCD  might be equivalent to a string theory. I shall  use the
exact solution of pure QCD, \ie with no quarks, as a testing ground for these
ideas. In particular I examine the free energy  of QCD  on an arbitrary
space-time background. Because of the simplicity of two dimensional gauge
theories, there being no propagating degrees of freedom, the partition function
can be  explicitly calculated on an arbitrary Euclidean manifold. I study the
\N expansion of the theory and find that many features of this expansion are
consistent with the string picture, indeed would be hard to understand
otherwise.

\chapter{\bf   Two Dimensional QCD  }

\section{\bf The Theory}

Two dimensional Quantum Chromodynamics (${\rm QCD}_2$) is the
perfect testing ground
for the idea that a confining gauge theories might be equivalent
to a
string theory. First, many features of the theory
are stringier in two than in four  dimensions.
For example, linear confinement is a perturbative
feature which is exact at all distances. Most important
is that the theory is exactly solvable.
This is essentially because in two dimensions
gluons have no physical, propagating
degrees of freedom, there being no transverse
dimensions. In fact \Q is the next best
thing to a topological field theory. The correlation
functions in this theory will depend, as we shall see,
only on the {\em topology } of the manifold on
which formulate the theory and on its {\em area}.
For this reason it is possible to  solve the theory
very easily and explicitly.

Consider for example the expectation value of
the Wilson loop for pure \Q,
$\Tr_R P e^{\oint_C  A_\m dx^{\m}}$,
for any contour,
$C$, which does not intersect itself. Choose
an axial gauge, say $A_1=0$, then the
Lagrangian is quadratic, given by $ \oh \Tr E^2$,
where $E = \partial_1 A_0$ is the electric field.
The Wilson loop describes  a pair of charged particles
propagating in time. This
source produces, in two dimensions, a constant electric field. The
Wilson loop is then given by the exponential
of the constant energy of the pair
integrated over space and time. This yields,
$$\Tr_R P e^{\oint_C  A_\m dx^{\m}}=
e^{- g^2 C_2(R) A} \sp ,\eqn\aa $$
where $g$ is the gauge coupling, $C_2(R)$
the quadratic Casimir operator for representation
$R$ and $A$ the area enclosed by the loop. The
expectation value of more complicated
Wilson loops that do self intersect can
also be calculated.  Kazakov and Kostov
worked out a set of rules for such loops
in the large N limit \refmark\kazkostwo. They are quite complicated.
\Q with quarks is also  soluble, at least in the
large $N$ limit. The meson spectrum was solved
for $N \to \infty$ by 't Hooft. It consists of an infinite
set of confined mesons with  masses $m_n$
that increase as $m_n^2 \sim n$. This provides one
with a quite realistic and very
instructive  model of quark confinement \refmark\thooft, \refmark\calgro .

Is \Q describable as a string theory? The
answer is not known, although there is
much evidence that the answer is yes.
I shall present below  some additional  evidence that supports this hypothesis.
To simplify matters I shall discard the quarks and consider the pure
gauge theory. This would correspond to a theory of closed strings only,
quarks are attached to the ends of open strings. We shall consider the
partition function for a $U(N)$ or $SU(N)$ gauge theory, on an arbitrary
Euclidean manifold ${\cal M}$,
$${\cal Z}_{\cal M} = \int[{\cal D} A^\m]
e^{- {1\over 4 g^2} \int_{\cal M} d^2 x\sqrt{g}
\Tr F^{\m \n}  F_{\m \n}}\sp . \eqn\bb $$
One might  think that in the absence of quarks the theory is totally
trivial, since in two dimensions there are no physical gluon degrees of
freedom. This is almost true, however the free energy of the gluons will
depend non-trivially on the manifold on which they live. In fact, one
cannot simply gauge the gluons away. If, for example, ${\cal M}$
contains a non-contractible loop $C$, then if  $\Tr  P
e^{\oint_C  A_\m dx^{\m}}\neq 1 $, one can not gauge $A_\m$ to zero
along $C$. Thus, the partition function will be sensitive to the {\em
topology} of  ${\cal M}$.

Although non-trivial the theory is extremely
simple, almost as simple as a topological theory. It is easy to see that
the partition function will only depend on the topology and on
the area of the manifold ${\cal M}$. This is because the theory is invariant
under all {\em area preserving diffeomorphisms}. To demonstrate this
note that the two-dimensional field strength can be written as
$ F_{\m \n}= \epsilon_{\m \n} f$, where $ \epsilon_{\m \n}$ is the
anti-symmetric tensor  and $f$ a scalar field. Thus the action is
$ S= \int \Tr f^2 d\m$, where $d\m = \sqrt{g} d^2x $ is the volume
form on the manifold.  This action is independent of the metric, except
insofar as it appears in the volume form. Therefore the theory is
invariant under area preserving diffeomorphisms ($W_{\infty}$). The
partition function can thus  only depend on the topology and on
the area of the manifold ${\cal M}$,
$${\cal Z}_{\cal M} = {\cal Z}[G, g^2, A, N]={\cal Z}[G, g^2 A, N] \sp ,
\eqn\cc$$
where $G$ is the genus of ${\cal M}$.

Now we can state the conjecture that the logarithm of this partition
function, the free energy, is identical to the partition function on some
string theory, with target space ${\cal M}$, where we would identify
the string coupling with \N and the string tension with $g^2 N$,
$$\ln {\cal Z}[G, g^2 A, N] = {\cal Z}_{  {\cal M}}^{\rm String} [g_{\rm
st}={1\over N}, \alpha'=g^2 N  ] \sp .
\eqn\dd $$

 As a candidate for the type of string theory I am thinking of consider
the Nambu action, wherein
$${\cal Z}^{\rm String}_{\cal M}= \sum_{h = \rm genus} (g_{\rm
st})^{2h-2} \int {\cal D} x^{\m}(\xi) e^{ \int d^2\xi \sqrt{g}} \sp ,
\eqn\dd $$
where $g$ is the determinant of the induced metric,
$$g= \det[g^{\alpha \beta}] = \det[ {\partial  x^\m \over
\partial \xi_{\alpha}}
{\partial  x^\n \over  \partial \xi_{\beta}} G_{\m \n}(x)]  \sp , \eqn\ee $$
and $G_{\m \n}(x)$ is the metric on the manifold ${\cal M}$.
This string theory, when the target space is two-dimensional, is indeed
invariant under area preserving diffeomorphisms of the target space.
To see this note that $\sqrt{g} = |{\partial  x^\m \over \partial
\xi^\alpha}| \sqrt{G}$, which is obviously unchanged by a map $x^\m
\to x'^\m$ as long as the Jacobian of the map, $   |{\partial  x^\m \over
\partial x^{'\n}}| =1$.
Actually the Polyakov action with a two-dimensional target space also has a
$W_{\infty}$ symmetry, although is is realized in a very nonlinear fashion. One
might speculate that this is related to the well known $W_{\infty}$ symmetry of
the $c=1$ string theory \refmark\gropol.  Unfortunately the only way we know to
quantize the Nambu action  is to transform it into the Polyakov action
\refmark\pol, which upon quantization yields the standard non-critical string.
This is not what we want to do here, since the resulting theory is not even
Lorentz invariant.

Is there another quantization of the Nambu string that differs from the
Polyakov quantization in two-dimensions? The answer is not known. However the
arguments that lead to the Polyakov action \refmark\pol~do require some
dynamical assumptions. They are especially suspect in low dimensions of the
embedding space, where the generic maps are singular.
In two dimensions the Nambu action would be topological were it not for these
singularities. The area of the surface mapped out by the string is only
non-trivial because of {\em folds}. If the Jacobian $  |{\partial  x^\m \over
\partial
\xi^\alpha}|$ never vanishes, then the Nambu action is equivalent to the
oriented area--\ie the topological {\em winding number}, $ n=\int{d^2\xi
\det\left({\partial  x^\m \over \partial
\xi^\alpha}\right)}$, which measures how many times the we cover the target
space. The true area differs from this by the contribution of folds. These
occur when $ \det\left({\partial  x^\m \over \partial
\xi^\alpha}\right)=0$, due to the vanishing of one of the eigenvalues of $
\left[  {\partial  x^\m \over \partial
\xi^\alpha}\right]$. They occur on smooth curves which terminate at Whitney
points. Such singularities cannot be seen in the usual conformal gauge, where
the induced metric is written as $ g_{\alpha\beta}=e^\phi \hat
g_{\alpha\beta}$,
where $ g_{\alpha\beta}$ is a fiducial, non-singular metric.

This remarks lead one to suspect   that the standard Liouville approach to  the
two-dimensional string might be modified. One approach might be to write the
action as
the topological winding number plus the contribution of the folds, and to
replace the latter by an action due to pointlike particles moving along
trajectories.  Thus, locally the Lagrangian is a total derivative  as long as
the determinant, $\det\left({\partial  x^\m \over \partial
\xi^\alpha}\right)$,
does not vanish. The total derivative yields the winding number of the map and
thus the action can be rewritten as
$$  S= \int{d^2\xi |\det \left( {\partial x \over \partial \xi} \right)}|
\sqrt{G}
= n + 2 \sum_i\oint_{C_i}dx^\mu \epsilon_{\mu\nu} x^\nu\sqrt{G} \sp ,
\eqn\eqqw
$$
where $C_i$ denotes a curve on which $ \det \left( {\partial x \over \partial
\xi} \right)=0$. The additional, non-topological term, counts the area of the
folds of the map. It is tempting to replace this theory of maps with a theory
of particles (fermions, perhaps, to give the correct counting of the loops),
moving along trajectories which correspond to the boundaries of the folds, with
some sort of (gauge?) interaction to account for the area term. However, to
date, we have not been able to construct such a particle action to replace the
Nambu action.

In the absence of a well defined candidate for the string action and measure we
shall proceed in the reverse,  experimentally constructive fashion. We shall
examine the partition function of QCD$_2$, expand it in a power series in
${1\over N}$,  determine whether this expansion can be interpreted  in terms of
maps of two dimensional surfaces onto two dimensional surfaces and finally,
hopefully, reconstruct the string action and measure.
\section{\bf  Evaluation of the Partition Function}

The partition function for \Q can easily be evaluated by means of the following
idea,  originally due to Migdal \refmark\migdal. The trick  is to use a
particular lattice regularization of the theory which is both exact and
additive. For the lattice we take an arbitrary triangulation of the manifold
(by triangulation I mean an arbitrary representation of the surface by means of
polygons),
as depicted in Fig.1,

\centerline{\epsffile{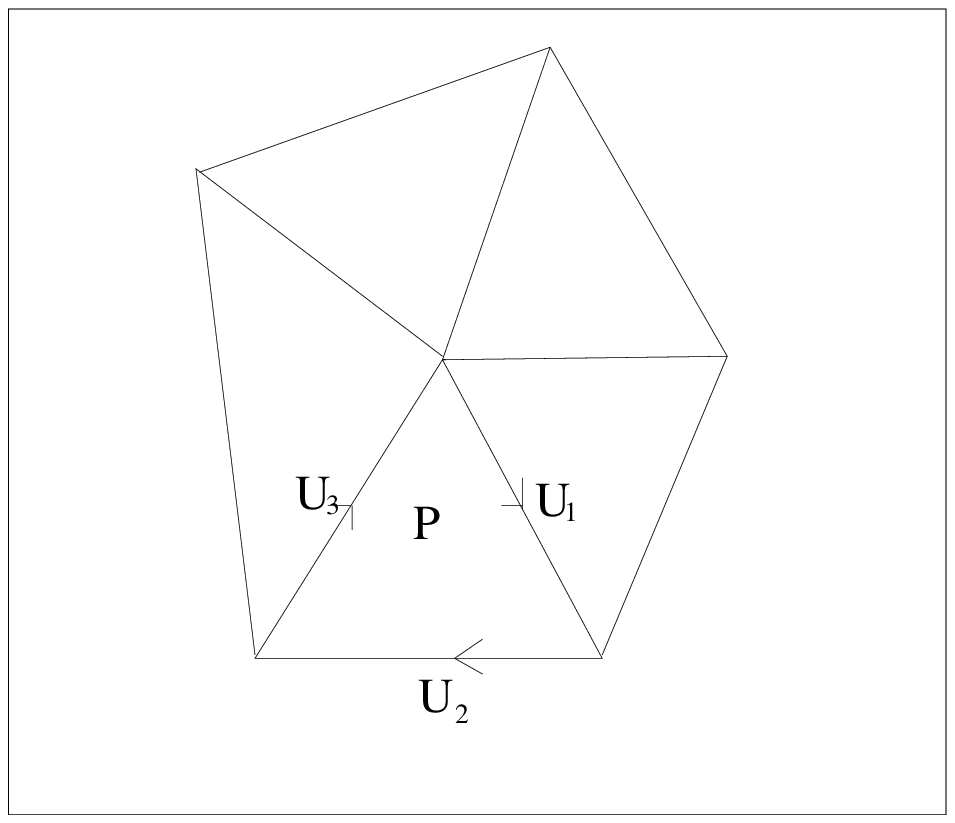}   }
\centerline{ Fig. 1  \hskip .1truein   A triangulation of ${\cal M}$ with group
matrices along the edges.}
\vskip .2 truein

\noindent and define the partition function as
$$ {\cal Z}_{\cal M} = \int \prod_L d U_L \prod_{\rm plaquettes} Z_P[U_P]
\sp, \eqn\ff $$
where $U_P= \prod_{L\in {\rm plaq.}} U_L$, and $Z_P[U_P]$ is some appropriate
lattice action. Any action will do as long as it reduces in the continuum limit
to the usual continuum action. Instead of the Wilson action, $Z_P(U)= e^{-{1
\over g^2}\Tr (U+U^{\dagger})}$, we shall choose the {\em heat kernel action},
$$ Z_P= \sum_R d_R  \, \chi_R(U_P) e^{-g^2 C_2(R) A_P } \sp ,\eqn\gg
$$
where the sum runs over representations $R$ of $ SU(N) $ (or $U(N)$),
$d_R $ is the dimension of $R$, $\chi_R(U_P)$ the character of $U_P$
in this representation, $C_2(R)$ the quadratic Casimir operator of $R$
and $A_P$ the area of the plaquette.

It is easy to see, using the completeness of the characters to expand about
$Z_P   \buildrel  {U_L \to 1 +i A_\m dx^\m}  \over\to
\sum_R d_R \chi_R(U_P)= \delta(U_P-1) +\dots  $, that in the continuum limit of
this theory reduces to ordinary Yang-Mills theory. What is special about the
heat kernel action is that it is additive. Namely, we can integrate over each
link on the triangulation and the action remains the same!  Thus we can
integrate out  $U_1$, which appears in precisely two triangles, as illustrated
in Fig. 2, using the

\centerline{\epsffile{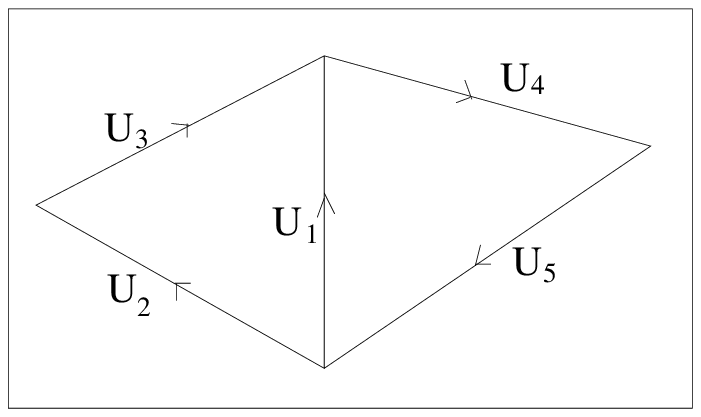}   }
\centerline{ Fig. 2  }
\vskip .1 truein
\noindent orthogonality of the characters,
 $\int dV \chi_a(XV) \chi_b(V^{\dagger} Y) ={\delta_{ab}\over
d_a} \chi_a(XY) $, and obtain,
$$\int dU_1 Z_{P_1}(U_2 U_3 U_1) Z_{P_2}(U_1^{\dagger} U_4
U_5)=  Z_{P_1+P_2}(U_2 U_3 U_4 U_5) \sp . \eqn\ii $$
This  formula expresses the  unitarity of the action, since in fact
$Z_P(U)= \langle U| e^{-g^2 A \Delta} |1\rangle  $, where $\Delta$ is the
Laplacian on the group.

We can use this remarkable property of the heat kernel action to argue that the
lattice representation is exact and  {\em independent of the triangulation}.
This is because we can use \ii~ in reverse to add as many triangles as desired,
thus
going to the continuum limit. On the other hand we can use \ii~ to reduce the
number of triangles to the bare minimum necessary to capture the topology of
${\cal M}_G$. A two-dimensional manifold of genus  $G$ can be described by a
$4G$-gon with identified sides: $a_1b_1a_1^{-1}b_1^{-1} \dots
a_Gb_Ga_G^{-1}b_G^{-1}$. The partition function can be written using this
triangulation as,
$${\cal Z}_{{\cal M}_G} = \sum_Rd_R e^{-{g^2} C_2(R) A} \int
\prod {\cal D} U_i {\cal D} V_i \chi_R[U_1V_1U_1^{\dagger}
V_1^{\dagger}\dots U_GV_GU_G^{\dagger}
V_G^{\dagger}] \,\, . \eqn\jj $$
We can now evaluate the partition function using the orthogonality of the
characters and the relation,
$\int {\cal D}U \chi_a[AUBU^{\dagger}] = {1\over d_a}  \chi_a[A]
\chi_a[B] $, to obtain \refmark\rus, \refmark\witetal,
$${\cal Z}_{{\cal M}_G} = \sum_R d_R^{2-2G}
e^{-{\lambda A\over N} C_2(R)  } \sp , \eqn\kk $$
where $\lambda \equiv g^2N$ is kept fixed.
Thus we have   an explicit expression for the partition function. It depends,
as expected, only on the genus and the area of the manifold.

So far we have considered both the groups $SU(N)$ and $U(N)$. In fact most of
the discussion given below will be the same for both of these groups.  It would
appear that they both yield expansions that have a stringy interpretation.  One
could also consider, using the same techniques,  other groups. It would be very
interesting to explore the group $O(N)$, which presumably would correspond to a
non-orientable string.
\section{ \bf The Large N expansion}

 The formula \kk~ for the partition function is quite complicated, being
written as a sum over all representations. The representations of $SU(N)$ or
$U(N)$ are labeled by the Young diagrams, with $m$ ($m \leq N$)  boxes of
length
$n_1 \geq n_2 \geq n_3 \geq \dots n_m\geq 0$. Such a representation  has
dimensions and Casimirs given by,
$$\eqalign{ C^{U(N)}_2(R)&=N \sum_{i=1}^m n_i +\sum_{i=1}^m n_i(n_i+1-2i) ; \cr
C^{SU(N)}_2(R)&=N \sum_{i=1}^m n_i +\sum_{i=1}^m n_i(n_i+1-2i) -{(\sum_{i}
{n_i})^2 \over N}; \cr
d_R &= {\Delta(h )\over \Delta(h^0)}, \sp h_i=N+n_i-i,\sp
h_i^0=N-i \cr \Delta(h) & =\prod_{1 \leq i<j\leq N} (h_i-h_j)   \sp . \cr}
\eqn\ll $$
 Thus we have a very explicit sum and one can, in principle, expand each term
in powers of \N and evaluate the sum.

Note that   the \N expansion is an expansion in powers of ${1 \over N^2}$, \ie
there are no odd powers of \N. This is because in the sum over representations
of
$SU(N)$ or $U(N)$, for every representation $R$, corresponding to a Young
tableaux, $ \lambda_R  $, there is a {\em conjugate} representation, $ {\bar
R}$, whose
Young tableaux, $ \lambda_{\bar R}    $, has its rows and columns interchanged.
For such conjugate representations
the dimensions and value of the Casimir operator satisfy (see  Appendix A.1),
$$\eqalign{{C^{U(N)}_2(R)\over N} &=  \sum_{i=1}^m n_i +
{\tilde C(R) \over N} \sp ;{C^{SU(N)}_2(R)\over N}  ={C^{U(N)}_2(R)\over N}
-{(\sum_{i} {n_i})^2 \over N^2}\cr
\tilde C(R) &= -\tilde C({\bar R}) \cr
d^2(R,N)& =d^2({\bar R} , -N)  \sp . \cr
}
\eqn\ewq$$
Now consider the \N expansion of \kk. Odd powers of \N can only arise from the
expansion of $d^2(R)e^{ -{\lambda A \tilde C(R) \over N} }$.  Summing over each
representation and its conjugate we get
$$   \sum_{k=0}^\infty{1\over k!}\left[ {\lambda A \tilde C(R)\over N}
\right]^k   \left( d^{2 -2G}(R,N)+(-1)^k d^{2 -2G}(
 R,-N) \right) \sp .\eqn\ewsd $$
But, since $d^{2 -2G}(R,N) = N^{n(2-2G) }\sum_{i} d_R^{G,i} {1\over N^i}  $, we
see that only even powers of  \N survive. This is as  expected for a matrix
model of Hermitean matrices--or a string theory of  closed orientable surfaces.

\section{\bf  Stringy Expectations}

What do we expect from the  \N expansion if the string conjecture is correct?
Consider the expansion in powers of \N of the free energy,
$$\ln[{\cal Z}_{{\cal M}_G}] = \sum_{g=0}^{\infty} {1\over N^{2g-2}}
f^G_g(\lambda A) \sp . \eqn\nn $$
\vskip .4 truein

\centerline{\epsffile{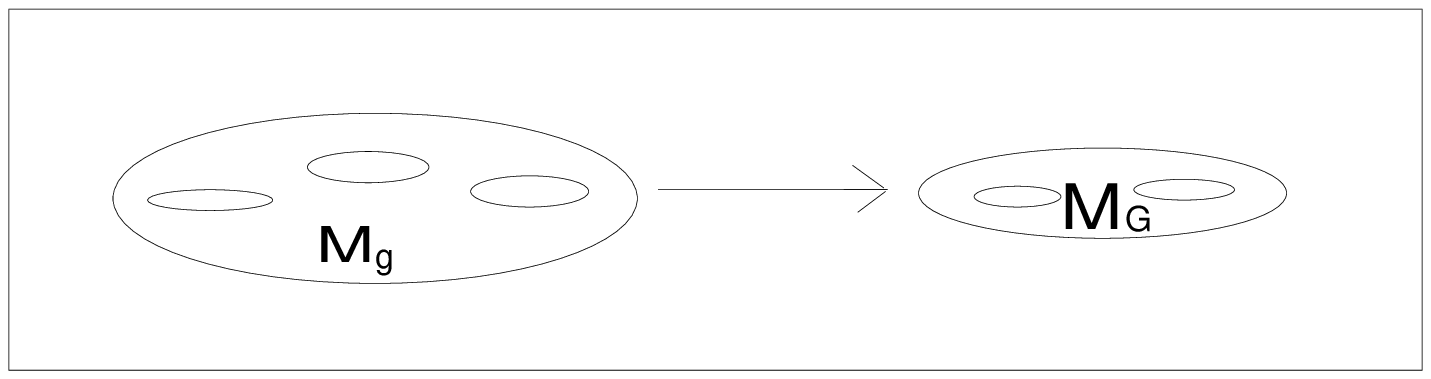}}
\centerline{ Fig. 3 \hskip  .3truein  A Map  from $ {\cal M}_g $  to   ${\cal M
}_G$     }
\vskip .3 truein

If this were given by a  sum over maps of a two-dimensional surface of genus
$g$, ${\cal M}_g$, onto a two-dimensional surface of dimension $G$, ${\cal
M}_G$, we would expect that $f^G_g(\lambda A) \sim ({1\over N})^{2g-2}
e^{-\lambda A n}$, where $n$ is the {\em winding number } of the map, \ie the
topological index that tells us how many times the map $x(\xi)$  covers ${\cal
M }$. This is the integral
of the Jacobian of the map $\xi \to x$, $\int d^2 \xi \det[{\partial x^\m\over
\partial \xi^i}]$, which differs from the Nambu area, $\int d^2 \xi
|\det[{\partial x^\m\over \partial \xi^i}]|$, since the surface can fold over
itself.

 However we can say more if the partition function is  a sum of smooth maps.
This is because   there is a minimum value that $g$ can take, given the genus
$G$ of the target space and the winding number $n$. Thus, for example, there
are no smooth maps of a sphere onto a torus,  or of a torus onto a genus two
surface. Similarly there are no smooth maps of a genus $g>1$ surface onto a
genus $g>1$ surface that wind around it more than once. To get an idea of the
relation between the genus of ${\cal M}_g$  and ${\cal M}_G$  consider
holomorphic maps, in which case the Riemann-Hurwitz theorem   states that
$2(g-1) =  2n(G-1) + B$, where $B$ is the total branching number.
In the case of  smooth maps there exists the following bound, known as {\em
Kneser's formula} \foot{ I thank A. Schwarz
for providing me with this reference.} \refmark\knesser ,
$$2(g-1) \geq 2n(G-1) \sp . \eqn\aaa$$

It is not difficult to picture the  maps that saturate this bound. These are
{\em covering} maps that do not have branchpoints or collapsed handles. To
picture these cut the genus $G$ surface, ${\cal M}_G$ along a cycle,
as in Fig.3; take $n$ copies of the resulting surface which has $G-1$ handles
and two boundary circles. Then join these  along the circles to form a surface
with $g=n(G-1) +1$ handles.
\vskip .35 truein

\centerline{\epsffile{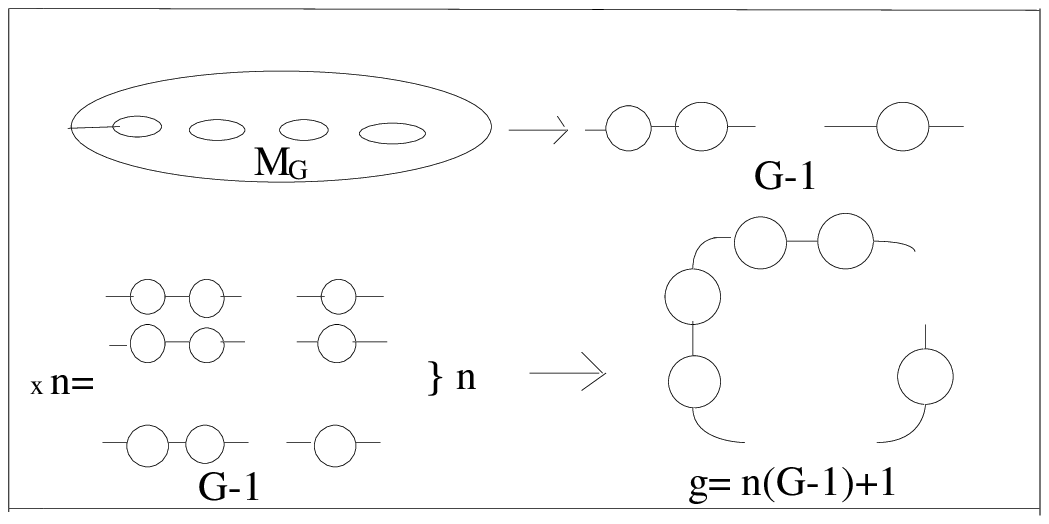}   }
\centerline{ Fig. 4  \hskip .1truein    A $n$-fold covering of  $ {\cal M}_G $
by  ${\cal M }_g$, where $g-1=n(G-1)$   }
\vskip .3 truein

If we add to ${\cal M }_g$ additional handles and map these to points
on ${\cal M }_G$ or if the map has branchpoints then clearly the Euler
character of ${\cal M }_g$ is increased and the bound improves.
The actual proof of Kneser's  formula is quite simple.   Essentially one shows
that  any map can be continuously deformed into a covering map with no branch
points or collapsed handles and in the process of deformation the Euler
character can only increase.

Thus, if \Q is described by a string theory we would expect that
$$ f^G_g(\lambda A) =  \sum_n \cases{ 0  & if $(g-1) < n(G-1)$ \cr
e^{-n\lambda A} \omega^n_{g,G}(A) & otherwise \cr} \sp .\eqn\mm $$
We might expect that the coefficients $\omega^n_{g,G}(A)$ are related to the
number of maps from ${\cal M}_g$ to ${\cal M}_G$ with winding number $n$. If
Kneser's formula is saturated, namely if $(g-1) =  n(G-1) $ we might conjecture
that $\omega^n_{g,G}(A)$ is determined topologically, it counts the number of
maps up to homotopy. Thus it should be independent of $A$.  However, when
$(g-1) >  n(G-1) $    then the map  must contain collapsed handles or
branchpoints. In that case we would expect to find powers of the area since the
branchpoints or collapsed handles can occur anywhere on the surface. We might
expect $\omega^n_{g,G}(A)$  to be a polynomial in $A$, whose maximum power is
the number of branchpoints plus collapsed handles.  Thus we would conjecture
that

$$ \omega^n_{g,G}(A) = \sum_{i=0}^{g-1-n(G-1)} \omega^{n,i}_{g,G}  A^i \sp .
\eqn\ewder
$$

We therefore have a stringent set of conjectures that we can use to test
whether \Q is describable as a theory of maps.

\chapter{\bf Comparison of \Q with the String Expectations}

To do  test our expectations, summarized by equations  \mm, \ewder, we must
expand \kk~ in a power series in \N. This we do below.

\section{\bf Genus $>1$ }

First, we shall consider the case of $G >1$.   For these manifolds the sum over
representations of the group converges very rapidly due to the factor
 $  \left(  {1 \over  d_R} \right)^{2G-2}$. In fact, the leading term in the
partition function is given by the fundamental representation, with
$   d =N, \sp C_2=N$, and  contributes
 $ \left(  {1 \over N} \right) ^{2G-2}e^{-\lambda A}$ to the partition
function. This clearly corresponds to the map of a genus $g=G$ surface onto our
genus G  manifold with winding number one.  As we might expect this appears
with weight one in $Z$ or in $\ln{Z}$.

To go beyond this we use the fact that for large $N$ the dimensions of the
representations behave as (see Appendix A.1),  $ d_{R_n} =  {f_R \over n!}N^n$,
for a representation whose Young tableaux has  $n$ boxes,
and $f_R$ is the dimension of the representation of the symmetric group of $n$
objects, $S_n$, corresponding to the same Young tableaux.
For fixed $n$ the Casimir operator of the representation grows as $ C_2(R_n)
\sim Nn$. Therefore the leading terms in the \N expansion are,
$$Z_{G} \to \sum_{n=0}^\infty ({1\over N})^{2n(G-1)}e^{-n\lambda A}
\sum_{r={\rm rep\,\, of\,} S_n} \bigl[{n!\over f_r}\bigr]^{2(G-1)} \sp ,
\eqn\rr $$
where the second sum is over representations of the symmetric group $S_n$.
The corrections to \rr~are of order \N (coming from the corrections to the
formula for the dimensions $d_R$), and of order $ {\lambda A \over N}$ (coming
from the corrections to the formula for the Casimir operator.)

This structure is precisely what we expected from the string picture. If we
consider the \N  expansion of   $ \ln{Z}$ we  obviously find that
$$\omega_{g,G}^n=0; \sp  {\rm  if}\sp    (g-1)<n(G-1)$$
and that the term which corresponds to the covering map,  for which
$g-1=n(G-1)$,  has a coefficient which is independent of the area. Furthermore,
since every power of $A$ in the  expansion is accompanied by a factor of \N, we
see that \ewder~ is satisfied. Thus all of the string expectations are true to
all orders in the \N expansion. This would be very hard to understand if the
string picture were not true.

 Recently we have been able to do more and (with W. Taylor) show  that
for the covering maps for which $g=1 +n(G-1)$,
$\omega^n_{g,G}   $  is precisely equal  to the number of topologically
inequivalent maps from the genus $g$ manifold onto the genus $G$ manifold with
winding number $n$\refmark\tobewat !

\section{\bf The Torus}

In the case of the torus, ($G=1$),  the sum is independent of the dimensions of
the representations,
$$ Z_{G=1}  =\sum_R  e^{-{\lambda A\over N}  C_2(R)}\sp . \eqn\torss
$$

 One can easily derive  (for $SU(N) $ or  $ U(N) $),
the leading terms as $ N\rightarrow\infty$,
$$\eqalign{Z_{G=1}& =\sum_R  e^{-{\lambda A\over N}  C_2(R)} \to
\!\!\!\!\!\!\sum^\infty_{n_1 \geq n_2  \geq \dots n_N \geq 0} \!\!\!\!\!\! e^{-
\lambda A \sum_in_i
}= \!\!\!\!\!\! \sum^\infty_{k_i =n_{i+1}-n_i \geq 0} \!\!\!\!\!\! e^{- \lambda
A \sum_n n k_n}    \cr
& = \prod_{n=1}^N {1 \over  1-e^{- n\lambda A  }} \to \exp \left[   N^0 \ln
\eta(e^{-\lambda A})\right]   \sp , \cr }
\eqn\qqpl
$$
where  $\eta(x)$ is the Dedekind function, $\eta(x)= \prod_{n=1}^\infty
(1-x^n)^{-1}$.
This is as expected.  These terms corresponds to maps of a torus   ($g=1$) onto
a torus ($G=1$)
with any winding number. The $   N^2 $ term,  which would  correspond  to the
sphere ($g=0$) is,  as expected, absent.

The coefficient of the expansion of the free energy  can easily be interpreted.
We see that,
$$\ln{ Z}= -N^0 \ln \eta(-e^{-\lambda A})= N^0\sum_n   e^{-n\lambda A} {1 \over
2n}\sum_{a\cdot b=n} (a+b)\sp .
\eqn\eqqwsa
$$
It is easy to see that the coefficient of $e^{-n\lambda A}$ counts, as
expected, the number of different maps of a torus onto a torus $n$ times
\refmark\tobewat (the above  sum runs over  maps for which the two
cycles  of the torus wind $a$ and $b$ times respectively around the two cycles
of the target space torus).

 The corrections to this leading term are, again, in accord with our stringy
expectations. The corrections appear in powers   of  $ {\lambda A \over N}$ (in
the case of $SU(N)$ there are also powers of $ {\lambda A \over N^2}$),
thus they  increase 16the genus of the internal space and presumably
corresponds to the occurrence of a new branchpoint, or collapsed handle.
Indeed the form of these corrections is very suggestive. For example, the
correction to $\ln Z$ of order $ {\lambda A \over N^2}$, (in the case of
$SU(N)$), is given by
$$  {\lambda A\over N^2}
\sum_{n=1}^\infty e^{-n \lambda A}[{n \over 2}\sum_{ab=n}(a+ b) +\sum_{ab+cd=n}
ac] \sp . \eqn\qq $$

The \N expansion is not  likely to be convergent. We expect the usual
asymptotic series with  a behavior like  $ \sum_k {1 \over N^k} k!$,
corresponding to the  expected essential singularity in \N of the form
$e^{-N}$. In \Q these would arise from instantons, whose action is proportional
to $ {1 \over g^2}={N \over  \lambda}$. In string theory these would come from
string instantons and be very interesting to understand.
These singularities can be seen from our expansion if we take more care. For
example, in the derivation of \qqpl~there are corrections that come from the
fact that the product has a cutoff at $n=N$. A more careful
calculation yields,

$$\eqalign{\ln{Z}&=  \ln{\eta(e^{-\lambda A})}
+\sum_{n=N+1}^\infty\ln{(1-e^{-n\lambda A})}  \cr & \sim
\ln{\eta(e^{-\lambda A})} -
{e^{-\lambda A(N+1)} \over (1- e^{-\lambda A})}
- \oh {e^{-\lambda A(2N+2)} \over (1- e^{-2 \lambda A})} + \dots
\sp . \cr}
\eqn\qqt
$$

\section{\bf The Sphere}

The hardest case  to expand in powers of \N is   that of the sphere ($G=0$),
since  the sum over representations blow up rapidly. It is not even evident
that there exists a tamed large $N$ expansion when we take the logarithm of
${\cal Z}$.  Here the inequalities described above are trivially satisfied, and
we get contributions from surfaces of all  genus. Nonetheless the structure of
the terms is of great interest.

For large $N$, $C(R_n) \buildrel N \to \infty \over \to N\sum_i n_i =Nn$,  it
would appear reasonable to break up the sum in \kk~ into a sum over
representations with $n$ boxes  in the Young  tableaux.
Thus,
$$ Z_{G=0} =\sum_R d_R^2 e^{-{\lambda A\over N} C_2(R) }
\buildrel N \rightarrow \infty \over  \longrightarrow   \sum_n \sum_{R_n}
d_{R_n}^2 e^{-n\lambda A}(1 +\dots)  \sp .\eqn\pp $$
To evaluate this we need to evaluate the   sum over the squares of the
dimensions of the representations of $SU(N)$ or $U(N)$,
 $ \sum_{R_n} d_{R_n}^2$.  As described in Appendix A.2, this can be done using
a method of discrete
 orthogonal polynomials, yielding,
 $$\sum_{R_n} d_{R_n}^2 =  {N^2 +n -1 \choose n } \sp .\eqn\ppp $$
Thus,  it follows that,
$$ Z_{G=0} \to \exp[-N^2 \ln(1-e^{-\lambda A})   +O(N^0) + \dots ]  \sp .
\eqn\pqp $$
 However, if we examine the sum in \pp~ we see that it is dominated by terms
with $n\sim N^2$. Thus the dominant representations have $n_i \sim N$.
Therefore the corrections to the $nN$ term in the Casimir operator, namely
$\sum_{i=1}^m n_i(n_i+1-2i)$ (plus , in the case of $SU(N)$, $  -{(\sum_{i}
{n_i})^2 \over N}$), are of equal importance.
One can extend the method of discrete orthogonal polynomials to deal with this
case, although the complete structure of even the leading (order
$N^2$), term is not known.  Some of the terms in this expansion are
given by (with $x= e^{-\lambda A}$),

$${\cal Z}_0= -N^2\left[  \ln(1-x)  + {2\lambda A \over(1-x)^2}+\dots  \right]
+ N^0 \left[  {\lambda^2 A^2x^2 \over (1-x)^4} +\dots\right]+\dots \sp .
\eqn\pqqp
$$

It is not difficult  to show that the behavior of the $N^2$ terms is
in accord with string expectations\refmark\tobewat. In particular the
coefficient of the term which behaves as
$e^{- n \lambda A}$, which should correspond to maps of the sphere onto the
sphere with winding number $n$, is proportional to   $A^{2k-2}$. This is what
we expect
since such a map should have at least $2k-2$ branchpoints, which can be located
anywhere on the surface.

\chapter{\bf Conclusions}

In this paper I have explored the possibility of representing \Q as a string
theory. The results of the \N expansion of the theory
 look  precisely like what we would expect from string considerations. What
remains to be understood are the all the rational numbers that appear as
coefficients of the  powers of $e^{-\lambda A}$ and of \N in terms of the
counting of maps of ${\cal M}_g$ onto ${\cal M}_G$. Some of these are
understood\refmark\tobewat, but not all.  Once we can reconstruct the QCD
partition function as a sum of maps it   remains to construct a string action
that reproduces these counting rules.  Then we would like to extend the program
to include Wilson loops and quarks into the theory. Finally we would like to
generalize the string representation to four dimensions. Clearly, much remains
to be done.

\appendix
\section {\bf  Conjugate Representations }

We shall consider the relation between the dimensions and quadratic Casimirs of
conjugate representations of $U(N)$ (or $SU(N)$), \ie between the
representation
 $R$, corresponding to a Young tableaux, $ \lambda_R  $,  and the   {\em
conjugate} representation, $ {\bar R}$, whose Young tableaux, $ \lambda_{\bar
R}    $, has its rows and columns interchanged.
Take the Young tableaux  $ \lambda_R  $ to have rows  of lengths $n_1  \geq
n_2 \geq n_3 \geq \dots \geq n_N  \geq 0 $. The conjugate Young tableaux,  $
\lambda_{\bar R}    $, will have rows of lengths  $m_1  \geq  m_2 \geq m_3 \geq
\dots \geq m_N \geq 0$. The relation between the $n_i$'s and the $m_i$'s is
given by
$$  m_k= \sum_i \theta (n_i+1-k) \sp ;  \sp  \theta(n) =\cases{
1& for $n>0$ \cr
0 & for $n\leq 0$}  \eqn\apa $$
The quadratic  Casimir is given by $C_2(R) \!\!= \!\! N\sum_in_i  +
\sum_i\left[ n_i^2 -n_i(2i-1) \right]$.
Now we note that $ \sum_{k=1}^N m_k^2 = \sum_{k=1}^N  \sum_{i,j=1}^N\theta
(n_i+1-k)\theta (n_j+1-k)= \sum_{i=1}^N\sum_{k=1}^N\theta (n_i+1-k)
+2 \sum_{1 \leq i<j \leq  N}^N\sum_{k=1}^N\theta (n_i+1-k)= \sum_{i=1}^N n_i
+2\sum_{1 \leq i<j \leq  N}^N n_j =  \sum_{i=1}^Nn_i(2i-1)$.

Therefore

$$\eqalign{C_2(R) &= N\sum_in_i  + \sum_i\left[ n_i^2 -m_i^2  \right] \cr
C_2(\bar R) &= N\sum_i m_i  + \sum_i\left[ m_i^2 -n_i^2  \right] \sp .  \cr
} \eqn\apb$$
In particular, with the definitions of Section 2.2, $\tilde C(R) = -\tilde
C({\bar R})$.

The dimensions of conjugate representations are also simply related.
 We recall Weyl's formula for the dimensions,
$$ d_R  = {\prod_{1\leq i<j \leq N}(h_i-h_j)\over \prod_{1\leq i<j\leq N}( i-
j)}  \sp ; \sp h_i=N+n_i-1 \sp.\eqn\apc  $$
Now assume that the Young tableaux of this  representation has $r$ rows (\ie $
n_{r+1}= 0, \dots , n_N=0$). We can separate out of this formula the dimension
$f_R$ of the representation of the symmetric group of $ n\equiv
\sum_{i=1}^{r}{n_i}$ objects, corresponding to the partition $ \left[ n_1\geq
n_2\geq  \dots \geq n_r\right]$,
$$ f_{\left[ n_1, n_2, \dots n_r\right]}=
n! {\prod_{1\leq i<j \leq r}(h_i-h_j)\over \prod_{1\leq i<j\leq r}( i- j)} \sp,
\eqn\apf$$
To derive
$$  d_R =  {f_R \over n!}
 \prod_{i=1}^{r}{{(N+n_i-i)!}\over (N-i)!}\sp .
\eqn\apg$$

This formula is well suited for a \N expansion. Note that
$
{(N+n_i-i)! \over (N-i)!}= N^{n_i}\prod_{k=1}^{n_i}{\left( 1+{k-i \over N}
\right)}\sp ,$  where the index $k$ ($i$) in this formula runs over the columns
(rows) of the tableaux. Thus we can write the dimension as,
$$ d_R =  {f_R N^n \over n!} \prod_{v}   \left(  1+{\Delta_v \over N} \right)
\sp ,
\eqn\aph$$
where the product rums over all the cells of the tableaux and $ \Delta_v$ is
defined for each cell to be the column index minus the row index.

Now, since the dimension of conjugate representations of the symmetric group
are equal and since under conjugation the rows and column are switched, we
immediately derive that
$$ d_{\bar R}( N) ={f_R N^n\over n!} \prod_{v}   \left(  1-{\Delta_v \over N}
\right) = d_{R}(-N) \sp .
\eqn\apj$$

\section{\bf   Discrete Orthogonal Polynomials}

We wish to evaluate the partition function for \Q on the sphere
$$ Z_{0} = \sum_R d_R^2 e^{- \lambda  A {C_2(R)\over N}}\sp . \eqn\qza   $$
This is the hardest sum to perform, since the dimensions of the representations
appear in the numerator. To  leading order in \N, the Casimir operator
behaves as $C_2(R) \to N \sum_in_i + \dots  = Nn +\dots $, where  the
representation is specified by a Young tableaux with rows $n_1 \geq n_2 \geq
\dots  \geq n_m$.  Naively, it would appear that we can replace $C_2$ by $nN$
to get the the leading term. Using \ll~for the dimensions of the
representations,  we can write the leading term as
$$ Z_0 = {1\over N!} \sum_{h_1, h_2 \dots , h_N =0}^{\infty} \bigl[
{\Delta(h_1,h_2, \dots , h_N)\over \Delta(0,1, \dots, N-1)}\bigr]^2 \sp  e^
{\lambda A\bigl[{N(N-1)\over 2} - \sum_i h_i\bigr] } \sp , \eqn\qzb   $$
where the ${1\over N!}$ accounts for the fact that the $h_i$ are not ordered.

The fact that the dimensions are given by a Van de Monde determinant,
$\Delta(h_1,h_2, \dots , h_N) = \prod_{i<j}(h_i-h_j)= \det[h_i^{j-1}]$, suggest
that we use the method of orthogonal polynomials, so successful in the recently
solved  matrix models.The new ingredient here is that the variables $h_i$ are
discrete \foot{ I am indebted to Andrei Matysin  for developeing the technique
of discrete orthogonal polynomials described here.}. We introduce a complete
set of orthogonal polynomials, $P_n(h) \sp , n= 0,1, \dots \infty $, orthogonal
with respect to the measure, $\sum_{h=0}^\infty e^{-\alpha h}$, \ie
$$  \eqalign{    P_n(h) & = h^n + f^1_n h^{n-1} + \dots + f^{n-1}_n h + f^n_n
\sp , n=0,1, \dots ,\infty  \cr  & \sum_{h=0}^\infty e^{-\alpha h} P_n(h)
P_m(h) = \delta_{nm} r_n \sp . \cr} \eqn\qzc   $$
We  can then express the  Van de Monde determinant in terms of the $P_n$'s,
$\Delta(h_1,h_2, \dots , h_N)= \det[P_i(h_j)]$, expand the product of
determinants in \qzb~and use the orthogonality of the $P_n$'s to derive that
(for $\alpha = \lambda A$),
$$  Z_0 = {e^{\lambda A\bigl[{N(N-1)\over 2}\bigr]}\over
\prod_{k=1}^{N-1}\bigl[k!\bigr]^2} \prod_{n=0}^{N-1} r_n  \sp . \eqn\qzd   $$

Thus we need to calculate the $r_n$'s as a function of $\alpha$. To do this
differentiate the sum in \qzc~, for $n=m$, with respect to $\alpha$; use the
fact that
$\sum_{h=0}^\infty e^{-\alpha h}
  {\partial  P_n(h) \over \partial \alpha} P_n(h)=0$, and that  $ h P_n(h) =
h^{n+1} +f^1_n h^n + \dots = P_{n+1}(h) +[f^1_n-f^1_{n+1}] P_n(h)+ \dots$, to
derive,
$$ {\partial r_n \over \partial \alpha} = -\sum_{h=0}^\infty e^{-\alpha h}
 h P_n(h) P_n(h)= [ f^1_{n+1}-f^1_n]r_n \sp .\eqn\qze   $$
Next  differentiate the sum in \qzc~, for $m=n+1$, with respect to $\alpha$,
to derive,
$$ 0 =  \sum_{h=0}^\infty e^{-\alpha h}\bigl[-h P_n(h) P_{n+1}(h)+P_n
{\partial P_{n+1} \over \partial \alpha} \bigr] = -r_{n+1} +{\partial f^1_{n
+1} \over \partial \alpha} r_n \sp . \eqn\qzf   $$

These equations can be used to construct the $r_n$'s and the $f^1_n$'s.
Define $R_n \equiv {r_n\over r_{n-1}}$, so that $r_n = \prod_{i=1}^n R_i r_0$.
Then, using \qzf~, $R_{n } = {\partial f^1_{n  } \over \partial \alpha} $,
and thus,
$$ R_{n+1}- R_n= {\partial [f^1_{n +1}-f^1_n ] \over \partial \alpha} =
{\partial^2  \ln r_n  \over \partial \alpha^2} = \sum_{i=1}^n{\partial^2  \ln
R_i  \over \partial \alpha^2} +{\partial^2  \ln r_0  \over \partial
\alpha^2}\sp . \eqn\qzg   $$
With the ansatz that $R_n=n^2R$, we then derive that $(2n+1)R=
n{\partial^2  \ln R  \over \partial \alpha^2}+{\partial^2  \ln r_0  \over
\partial \alpha^2}$, which leads to,
$$ R = \oh {\partial^2  \ln R \over \partial \alpha^2}\sp; \sp R= {\partial^2
\ln r_0  \over \partial \alpha^2}\sp . \eqn\qzh   $$
Now $r_0=  \sum_{h=0}^\infty e^{-\alpha h} ={1\over (1-e^{-\alpha  })}$, and
thus
$$ R= {e^{-\alpha  }\over (1-e^{-\alpha  })^2} \sp ; \sp r_n = (n!)^2
{e^{-\alpha n }\over (1-e^{-\alpha  })^{2n+1}} \sp . \eqn\qzi   $$
Finally,
$$Z_0 = { 1 \over (1-e^{-\lambda A  })^{N^2}} = \sum_{n=0}^\infty
 e^{- \lambda A n} { N^2 + n -1 \choose n }  \sp . \eqn\qzk   $$

However, it is now clear that our initial assumption was incorrect. As noted in
the text the {\em corrections} to $C_2$ are not small as $N\to \infty$,  since
the sum in \qzk~is dominated by terms where $n\sim N^2$. The method of discrete
orthogonal polynomials can be extended to include these corrections, however
the resulting equations have so far proved intractable.
\vskip 1 truein
{\bf  Acknowledgements}

I thank  A. Matysin, A. Polyakov, E. Witten,  A. Schwarz, and W. Taylor for
helpful conversations.

\refout

\end